\documentstyle [12pt]{article}
\renewcommand{\baselinestretch}{2}

\title{Anomalous Cross Section from \\ Perturbation Theory }

\author{Taekoon Lee
        \\
        \\
        Fermi National Accelerator Laboratory\\
                P.O. Box 500, Batavia, IL 60510}

\date{}
\begin{document}
\maketitle
\begin{abstract}

We present the anomalous cross section for baryon number
violation in
the standard model from the perturbation of large-order
behavior of
forward scattering amplitudes to the order
$\left(\epsilon /n\right)^{8/3}
\ln\left(\epsilon /n\right)$.
An improved high energy behavior of the anomalous cross section
is observed. We also argue that
the asymptotic form of
$F(\epsilon  g) \equiv -g \ln \sigma_{ano} $ is
given in the form:
$F(\epsilon  g) \rightarrow d + c \cdot \epsilon  g $  for
$ \epsilon g \rightarrow \infty$ with $c, d $ constants
satisfying $c, d \geq 0$, and  $F(\epsilon g) >0$
for all energies. The constants are not determined.

\end{abstract}

\def\thepage{FERMILAB-PUB-94/385-T}
\thispagestyle{myheadings}
\newpage
\renewcommand{\baselinestretch}{2}
\pagenumbering{arabic}
\addtocounter{page}{0}

\newcommand{\be}{\begin{equation}}
\newcommand{\ee}{\end{equation}}
\newcommand{\bear}{\begin{eqnarray}}
\newcommand{\eear}{\end{eqnarray}}

\section{ Introduction}

We have recently shown that the large-order behavior of
Green's functions
has generally nontrivial energy dependence arising from the
Espinosa-Ringwald
type anomalous cross section, and in turn the anomalous
cross section itself
can be deduced from the large-order behavior of forward
scattering amplitudes \cite{R1}.  It was an extension of the
observation that in the
double-well potential problem in quantum mechanics, the
vacuum transition
rate determines the large-order behavior of the vacuum
to vacuum transition
function (here vacuum means one of the perturbative vacua),
and in turn the minimum element of the perturbative
series of
the latter reproduces the exponential part of the
former which is the imaginary part of the
vacuum transition function.  Recall that all the bubble
diagrams
for the
vacuum transition function are real and its imaginary
part arises
nonperturbatively from the instanton interactions.
We also proposed
calculating the anomalous cross section by doing perturbation
of the
Borel transform  of the forward scattering amplitudes about
its instanton-
anti-instanton singularity. We elaborate this proposition
further in this
paper, and clarify the relation between our
proposed method and the
energy expansion method in powers of $
\left( E/E_{0}\right)^{2/3}$ \cite{R2}.

In section 2 we review briefly the relation between large-order
behavior
and the anomalous cross section, and in section 3 we give a new
formulation
of our proposition in terms of $ \epsilon/n$ expansion,
where $n$ is the order
of perturbation. We calculate in section 4 the anomalous
cross sections from the $\epsilon/n$
expansion, and compare them to
those from the
energy expansion. Using the formalism in section 3, we argue in
section 5
that asymptotically $F(\epsilon  g) \equiv -g \ln
\sigma_{ano} $ either grows linearly in $ \epsilon g $ or
approaches to a nonnegative  constant. We
also show that $F( \epsilon g) > 0$
for all energies, and make a speculation on the value of
$d$ that appears in the asymptotic form.

\section{ Large-order behavior and anomalous cross section}

In the weak coupling limit of the standard model of weak
interactions
the anomalous cross section for baryon number violation
in two-body
scattering is given by \cite{R2}
\be
\sigma_{ano} \sim \left(\frac{1}{g}\right)^{\nu} e^{
-\frac{1}{g}
F_{0}(
\epsilon g)},
\ee
where
\be
F_{0}(\epsilon g) = 1 - U( \epsilon g).
\ee
$U(x)$ is given in the form \cite{R3},
\be
U(x) = \frac{1}{2} \left(3 x\right)^{\frac{4}{3}} -
\frac{1}{6} \left(3 x\right)^{2}- \frac{\lambda}{54}
\left(3 x\right)^
{\frac{8}{3}} \ln\left(3 x\right) + O\left( (3 x)^{
\frac{8}{3}}\right),
\label{r3}
\ee
with
\be
\lambda = 4 - 3 \frac{ m_{h}^{2}}{m_{w}^{2}},
\ee
where $m_{h}, m_{w} $ are the Higgs and gauge boson
 mass respectively.
The $\epsilon, g$ are defined as
\be
\epsilon = \frac{E}{m_{w}}, \hspace{.25 in} \mbox{ and}
\hspace{.25in}  g = \frac{\alpha_{w}}
{4 \pi},
\ee
with $ E,\alpha_{w} $ the c.m. energy  and the weak coupling
respectively.
$m_{w}$ is assumed to be independent of the weak coupling, and
$\nu$ is
a Green function dependent constant of order one. The weak
coupling limit
is to be understood as:
\be
g \rightarrow 0 \hspace{.25 in} \mbox{while}\hspace{.25in}
 \epsilon g \hspace{.25 in} \mbox{fixed},
\ee
and we always assume this limit in this paper.

Crutchfield \cite{R4} has shown that the
large-order behavior --- with
renormalon effects
not included --- of a Green's function can be calculated by doing
perturbation of its Borel transform around the
instanton-anti-instanton singularity. The energy dependence of
the Borel
transform of a forward scattering amplitudes arises from
instanton-
induced amplitudes   such as those in which
 the initial state particles are
attached to
the anti-instanton and the final states particles to the instanton,
or vice versa,
in an
instanton-anti-instanton background \cite{R5}.  This kind of
amplitudes gives
precisely the Espinosa-Ringwald type anomalous cross section
\cite{R8}.
Thus we are
interested in the Borel transform  of the anomalous cross section
\bear
\tilde{\sigma}(b)
   & = & \frac{1}{2\pi i} \int _{a -i \infty}^{a + i \infty}
   \sigma_{ano}(
g)\,  e^{ \left(\frac{b}{g}\right)} \,d \left( \frac{1}{g}\right)
 \nonumber \\
   & \sim & \int \exp\left( - z ( 1- b) + z U\left(
\frac{\epsilon}{z}
   \right)\right)
   z ^{\nu} d z
   \label{r7}
   \eear
with $  z \equiv 1/g$.
$\tilde{\sigma}(b)$ has a branch-point type singularity at
$ b = 1$ \cite{R6},
\be
  \tilde{\sigma}(b)  \rightarrow ( 1-b)^{ -(\nu + 1)}
  \hspace{.25 in} \mbox{ for} \hspace{.2in} b
  \rightarrow 1.
  \label{r8}
  \ee
With the perturbative  coefficients of forward scattering
amplitudes $ A $  defined  by
\be
A ( E, g) = \sum a_{n} g^{n},
\ee
the coefficients induced by the anomalous cross section
are given by
\be
a_{n} = (n-1)! c_{n-1}
\label{r10}
\ee
where $c_{n}$ is defined by
\be
\tilde{\sigma}(b) = \sum_{n} c_{n}\, b^{n}.
\label{r11}
\ee
 {}From Eq. (\ref{r7}), (\ref{r10}), and (\ref{r11}), we have
\be
a_{n} \sim \int e^{ -z \left( 1 - U\left( \frac{
\epsilon}{z}\right)
\right)} z^{N} d z
\label{r12}
\ee
with $ N \equiv n -1 + \nu$.

For large $N$, (\ref{r12}) may be evaluated in the saddle point
approximation to give
\be
a_{n}  \sim {\cal F}'' \left( z(N), N \right) ^{-\frac{1}{2}}
e^{ - {\cal F} \left( z(N),
N \right) }
\label{r13}
\ee
with
\be
{\cal F}( z, N) = z  - K(z) - N \ln z
\ee
and the saddle point $ z(N) $ satisfying
\be
\frac{ N}{z(N)} - 1 + \left. \frac{ d}{ d z} K(z)
\right|_{ z = z(N)} = 0,
\label{r15}
\ee
where
\be
K(z) = z\,U \left(
\frac{ \epsilon}{z}\right).
\ee
The minimum element of the perturbative series is
then given by
\be
a_{\bar{N}} g^{\bar{N}}
\sim e^{ - {\cal F} \left( z \left( \bar{N} \right),
\bar{N} \right)
+ \bar{N} \ln g }
\label{r16}
\ee
with $ \bar{N} $ satisfying
\be
   \left.\frac{ d}{d N} \left( -{\cal F }\left( z(N),
   N  \right) + N
  \ln g \right)
  \right|_{ N=\bar{N}} =\ln \left( z( \bar{N}) g \right) = 0,
  \label{r17} \ee
and using (\ref{r15}).
 {}From (\ref{r17})
\be
z( \bar{N} ) = \frac{1}{g}.
\label{r18}
\ee
Substituting (\ref{r18}) into (\ref{r16}),
\be
 - g \ln \left( a_{ \bar{N}} g^{ \bar{N} } \right) = 1 - U(
 \epsilon g) = F_{0} ( \epsilon g).
 \label{r19}
 \ee
 The minimum element of the perturbative series correctly
 produces the
 exponential part of the anomalous cross section. It is very
 important
 to note that in the weak coupling limit , the corrections
 to the saddle point
 approximation and the pre-exponential Gaussian determinant
 in (\ref{r13}) generate
 negligible terms, and thus Eq.(\ref{r19}) is exact.
The details can be
 found in the
 appendix.

\section{ $\frac{ \epsilon}{n}$ expansion}

Let us now see  how we can formulate the problem of the
anomalous cross section  starting from the following two
conditions: that the anomalous cross section $F$ can be
deduced by taking the minimum element of the perturbative
series of forward scattering amplitudes, and that $F$ is a
function of $ \epsilon g$ only. The latter condition was shown
to be true for the final state  corrections, but not
completely for the corrections related to the initial
state \cite{R7}.

Noting from (\ref{r12}) that $ (n-1)! c_{n-1}$ depends only on
$N$, we define a function $ \tilde{C}(N,\epsilon )$ by
\be
(n-1)! c_{n-1} \equiv N! \tilde{C}(N, \epsilon ).
\ee
 {}From the Borel singularity in (\ref{r8}) we find
\be
\tilde{C}(\infty, \epsilon ) = 1.
\ee
Now
\be
a_{n} g^{n} = (n-1)! c_{n-1} g^{n} \equiv e^{ - \frac{1}{g}
F(N, g, \epsilon )}
\ee
where
\be
 F(N, g, \epsilon ) = g N - g N \ln g N - g \ln \tilde{C}
 (N, \epsilon )
 \label{r23}
 \ee
 in the weak coupling limit and using the Stirling's
 formula.
 The maximum of $F(N, g, \epsilon )$ occurs at $ N= \bar{N}$
 satisfying
 \be
 g \bar{N}  = e^{ - H( \bar{N}, \epsilon )}
 \label{r24}
 \ee
 where
 \be
 H( N, \epsilon ) \equiv \frac{ \partial}{\partial N} \ln
 \tilde{C}(N, \epsilon ).
 \ee
 At $N = \bar{N}$,
 \bear
 F( \bar{N}, g, \epsilon ) &=& g \bar{N} -  g \bar{N} \ln g
 \bar{N} - g \int_{\infty}^{ \bar{N}}
 H( N, \epsilon ) d N    \nonumber \\
                &=& g \bar{N} -  g \bar{N} \ln g \bar{N} -
                \int_{\infty}^{g \bar{N}}
                H( N, \epsilon ) d (g N)
                \label{r26}
\eear
For $ F( \bar{N}, g, \epsilon )$ to be a function of $
\epsilon g$ only,
we see from (\ref{r24}) and (\ref{r26})  that $ H( N,
\epsilon )$ must be a
function of $\epsilon/N$ only, that is,
\be
 H( N, \epsilon ) = H \left( \frac{\epsilon}{N} \right).
 \ee
 Defining  new variables $ y_{0}, y$
 \be
 y_{0} = \frac{ \epsilon }{ \bar{N}}, \hspace{.25in} y =
 \frac{ \epsilon }{ N}
 \ee
 we can write  (\ref{r24}), (\ref{r26}) as
 \be
    \epsilon g = y_{0}\, e^{ - H(y_{0})}
   \label{r29}  \ee
    and
 \be
 F( \epsilon g) =  e^{ - H(y_{0})} \left( 1 +   H(y_{0} )
 \right) + \epsilon g
 \int_{ 0}^{ y_{0}}
 \frac{ H(y)}{y^{2}} d y.
 \label{r30}
 \ee
 Eq. (\ref{r29}), (\ref{r30}) are our main equations.
 In our formalism, all the
 information  on the anomalous cross section is contained in
 $H(y)$, and a natural perturbation scheme emerges, namely,
 the expansion of $ H(y)$  at $y = 0$  in powers of $y$.
 Since the integral term in (\ref{r30}) is heavily weighted toward
 $ y=0$, we expect it to be a good approximation scheme.
 For the consistency of Eq. (\ref{r30}), we note that
 \be
 \lim_{y \rightarrow 0} \frac{H(y)}{y} = 0
\label{r01}
\ee
should be satisfied.
 We emphasize that this $ \epsilon/n$ expansion is
different from the energy expansion;
 In our scheme the latter is an
 intermediate step to find the function $H$.

\section{ $F\left(\epsilon g\right)
 $from $\frac{ \epsilon }{n}$ expansion}

In this section, we calculate $ H(y)$ perturbatively
using the anomalous cross section from the energy expansion
in one-instanton sector, and then compute $F$ using
Eq. (\ref{r29}), (\ref{r30}).
Since we find it is instructive to calculate $ H$ and $F$
order by order, we present the calculations to each order
up to $ (\epsilon/n)^{8/3} \ln ( \epsilon/n)$.

There are two ways to calculate $H(y)$. The first one is
to expand the Borel transform $\tilde {\sigma}(b)$  in (\ref{r7})
around the instanton-anti-instanton singularity.
Expanding $ \exp(K(z))$ in the integrand in (\ref{r7})
in powers of $1/z$,
and then performing the $z$-integration exactly, $
\tilde {\sigma}(b)$ for a given $ U(x)$
can be calculated in power series of
$(1-b)$.
Once we have $\tilde {\sigma}(b)$, it is straightforward to
find $c_{n}$  by expanding $\tilde {\sigma}(b)$ about $b=0$.
A simpler way to find $H(y)$ is from the saddle point
approximation in (\ref{r13})---(\ref{r15}).
Since as shown in the appendix, the
saddle point approximation is exact in the weak coupling
limit, we have
\bear
\ln\left\{ ( n-1)! c_{n-1}\right\}
&=& \ln N! + \ln \tilde{C}(N, \epsilon )
\nonumber \\
                    &=& N \ln z(N) - z(N) + K\left( z(N)\right)
                    - \frac{1}{2} \ln {\cal F''}(z(N), N)
                    \nonumber \\
                    &=& N \ln N - N + \frac{1}{2} \ln N  +
                    N \ln \left( \frac{z(N)}{N} \right) +
                    K\left( z(N)\right) -z(N)
                    K'\left(z(N)\right)  \nonumber \\
                    &=& \ln N!  + N \ln \left( \frac{z(N)}{N}
\right)
                    + K\left( z(N)\right) -z(N)
                    K'\left(z(N)\right)
\eear
  using the Stirling's formula and
  the fact that $ {\cal F''} = 1/N$ in the
 weak coupling limit.
 Then
 \be
 \ln \tilde{C}( N, \epsilon ) = - N \ln \left( 1-
 K'\left(z(N)\right) \right)
                    + K\left( z(N)\right) -z(N)
                    K'\left(z(N)\right).
\label{r33}
 \ee
Taking the derivative of (\ref{r33}) in $N$, we find
\be
H( N, \epsilon ) = \frac{ \partial}{ \partial N} \ln
\tilde{C}( N, \epsilon ) =- \ln \left( 1-
K'\left(z(N)\right) \right) =
\ln \left( \frac{ z(N)}{N} \right).
\label{r34}
\ee
Solving $ z(N)$ in (\ref{r15}) perturbatively in $1/N$ with
$ U(x)$ given in (\ref{r3}), we find
\be
\frac{ z(N)}{N} = 1 - \frac{1}{6} \left(\frac{3 \epsilon }{N}
\right)^{\frac{4}{3}} +
\frac{1}{6} \left(\frac{3 \epsilon }{N} \right)^{2} +
\frac{5 \lambda }{162} \left(\frac{3 \epsilon }{N}
\right)^{\frac{8}{3}} \ln \left( \frac{3 \epsilon }{N} \right)+
O\left(\left( \frac{3 \epsilon }{N} \right)^{
 \frac{8}{3}}\right),
\ee
and
\be
H(y) = - \frac{1}{6} \left(3 y\right)^{\frac{4}{3}} +
\frac{1}{6} \left(3 y\right)^{2} +\frac{5 \lambda}{162}
\left(3 y\right)^
{\frac{8}{3}} \ln\left(3 y\right) + O\left( (3 y)^{
\frac{8}{3}}\right).
\label{r36}
\ee
We note that the equivalency of the two methods has been
explicitly checked to the order $ (3 y )^{8/3}$ using the
leading term of $U(x)$, i.e.,
\be
U(x) =    \frac{1}{2} \left(3 x\right)^{\frac{4}{3}}.
\label{r37}
\ee
Note that $H(y)$ in (\ref{r36}) satisfy Eq. (\ref{r01}).
This may explain why the leading term of $U(x)$ has a
power larger than unity. Eq. (\ref{r01}), for example,
does not allow a term of order $ (3 x)^{2/3}$.
We now compute $F$ order by order up to $\left(3 y
\right)^{\frac{8}{3}} \ln\left(3 y\right)$.

\subsection{ Leading order}

For this case, $ U(x)$ is given by (\ref{r37}) and
\be
H(y) = -\frac{1}{6} \left(3 y\right)^{\frac{4}{3}}.
\label{r38}
\ee
With (\ref{r38}), Eq. (\ref{r29}) is solvable for all energies,
and there is a
unique solution for a given $ \epsilon g$.
Solving (\ref{r29}) numerically, we plot $F$ in Fig.1. Note
that at low energies $ F $ and $F_{0}$ matches very well,
but at high energies there is a sizable difference between them.
$F$ gives a better high energy behavior. One may wonder
why $ F$ and $F_{0}$ are different in view of the discussions
in section 2. The reason is that $H(y)$ in (\ref{r38})
 is only part of
the series expansion of that defined in (\ref{r34}). If we had
solved $ z(N)$ in (\ref{r15}) and expanded  $H(y)$
in (\ref{r34}) to an
infinite order, the resulting $F$ would have been identical
to $F_{0}$.  In our formalism, there is no point
of expanding $H(y)$ to an order higher than that of $U(x)$.

\subsection{ Second order }

$U(x)$ to the second order is given by
\be
U(x) = \frac{1}{2} \left(3 x\right)^{\frac{4}{3}} -
\frac{1}{6} \left(3 x\right)^{2},
\ee
and
\be
H(y) = - \frac{1}{6} \left(3 y\right)^{\frac{4}{3}} +
\frac{1}{6} \left(3 y\right)^{2}.
\ee
The function $ y \exp (- H(y))$ is plotted in Fig.2,
and we see that
there is no solution for Eq. (\ref{r29}) for $ \epsilon g \geq  E_{o},
$ where
$E_{o} =0.53$. $E_{o}$ is the upper limit for the applicability
of our formalism  up to the second order. At energies
below $E_{o}$
there are two solutions for a given $ \epsilon g $. By integrating
$H(y)$ to obtain $ \ln \tilde{C}(N, \epsilon )$, it is easy to
check that the solution close to the origin is for the true maximum
of $F(N, g, \epsilon ) $ in (\ref{r23}),  and the other solution is
an artifact
of low order expansion of $H$. The $ F, F_{0}$ are  given
in Fig.3.

\subsection{ Third order}

$U(x)$ and $H(y)$ are given in (\ref{r3}), (\ref{r36}) respectively.
The functions $H(y),  y \exp (- H(y))$ are plotted in Fig.4 for
$\lambda = -1 $ and $ -0.6, $ and
we see that  Eq. (\ref{r29}) is solvable for all energies.
We plot $ g \bar{N}  $ and $ F, F_{0}$ in Fig.5.
 Note the large difference between $F$ and
$F_{0}$ at high energies for $ \lambda =-0.6$.
For $\lambda = -1$, $ g \bar{N}  $ oscillates around
unity. It can be checked that for $ 0 <\lambda \leq 0.5, $
$g \bar{N}( \epsilon g) $  is discontinuous, and $F$
not analytic---
though continuous---at the discontinuity.
For example, with $\lambda = -0.4$
 $g \bar{N}$ is discontinuous at $\epsilon g \approx 0.53$, and $F$
is not analytic at the energy.
For the reasons discussed in next section, this discontinuity
is believed to be  an artifact of the expansion
of $H(y)$ to this particular order.

\section{ Constraints on $U(x)$, $H(y)$}

In this section we study various constraints on the
functions $U(x)$, $H(y)$.
 {}From Eq. (\ref{r34}),
we see that for a given $U(x)$, $H(y)$ can be
written as
\be
H(y) = \ln y - \ln x(y)
\label{r02}
\ee
with $x(y)$ defined implicitly through the relation
\be
y = \frac{ x}{ 1 - U(x) + x U'(x)}.
\label{r03}
\ee
Since $\tilde{C}(N,\epsilon)$  is believed to be a smooth,
well-defined function, we also expect  $H(y)$ to be analytic
over the positive real axis. For $H(y)$ to be smooth and
well-defined for $y>0$, the r.h.s. of (\ref{r03}) should
be a monotonically increasing function in $x$, and
\be
   \lim_{ x\rightarrow \infty}
   \frac{ x}{ 1 - U(x) + x U'(x)}   = \infty.
   \ee
Since (\ref{r03}) should be  invertible,
 we also have a constraint on
 $U(x)$,
\be
1 - U(x) + x U'(x) > 0 \hspace{.25in}
\mbox{ for} \hspace{.25in} x > 0.
\label{r04}
\ee
 With the transformation rule given in (\ref{r02})
 and  (\ref{r03}), $H(y)$  may be thought as a dual function
 of $U(x)$.  It can be shown without difficulty that
 Eq. (\ref{r30}) with (\ref{r29}) is indeed the inverse transform
 of that defined in  (\ref{r02}), (\ref{r03}).

Now we note that
Eq. (\ref{r29}), (\ref{r30}) are  very suggestive of the
following asymptotic
form for$F$:
\be
F = const. + c \cdot \epsilon g
\ee
with
\be
c = \int_{0}^{\infty} \frac{H(y)}{y^{2}} dy.
\label{r42}
\ee
Using unitarity, we show that this is indeed the case
if $H(y)$ has a smooth asymptotic limit,
either finite or infinite.
This means  that $g \bar{N}$  also
has a smooth asymptotic limit, as can be seen from (\ref{r24}).
Now unitarity applied to Eq. (\ref{r30}) prohibits $H(y)$ from
approaching to $-\infty$ asymptotically,
and thus the asymptotic limit of
$g \bar{N}$ should be finite.
Now note that $x(y)$ defined in (\ref{r03}) is  a
monotonically increasing function and thus so is
\be
\ln y - H(y).
\ee
This, combined with the unitary condition on the asymptotic limit
of $H(y)$ mentioned above, implies that   the integral in
 (\ref{r42}) is rapidly convergent in the asymptotic region
and thus $c$ is finite.

Let  us now consider the integral term in (\ref{r30}),
\be
I( \epsilon g) \equiv \int_{0}^{y_{0}( \epsilon g)}
\frac{ H(y)}{ y^{2}} d y
\ee
with
$y_{0}( \epsilon g)$ defined through  (\ref{r29}).
Expanding $ I$ at $ \epsilon g=\infty$, we have
\be
I( \epsilon g) = c + d' \cdot \frac{ 1}{ \epsilon g} + \cdots
\label{r45}
\ee
where
\be
d' = - \lim_{y \rightarrow \infty} \frac{ e^{- H(y)}
H(y)}{1- y H'(y)}.
\ee
Substituting  (\ref{r45}) into (\ref{r30}), we find the
asymptotic form for $F$:
\be
F \rightarrow d + c \cdot \epsilon g
\ee
with
\be
d =  \lim_{y \rightarrow \infty} \left(  e^{- H(y)}
\left( 1 + H(y) \right) - \frac{ e^{- H(y)}
H(y)}{1- y H'(y)} \right).
\label{r50}
\ee
Note that unitarity requires
\be
c \geq 0. \label{r49}
\ee
When $c=0$, $F$ approaches to a constant.
Substituting the asymptotic form of $U(x)$,
\be
U(x) \rightarrow 1 - d - c \, x
\ee
into (\ref{r04}), we find
\be
d \geq 0.
\ee

Let us now show that
\be
 F(x) >0 \hspace{.25in} \mbox{for} \hspace{.25in}
 x> 0.
\label{r06}
\ee
First note that
\be
F(0) = 1, \hspace{.5in} F(\infty) \geq 0.
\ee
To prove (\ref{r06}), let us suppose that at some finite value
of $ x$,
$F(x) \leq 0$. Then $F(x)$ must have a minimum $F(x_{0})$ at a
finite $x_{0}$ satisfying
\be
F(x_{0}) \leq 0.
\ee
However this would contradict (\ref{r04}), because
\be
U'(x_{0}) = 0.
\ee
Thus the conclusion in (\ref{r06}) follows.
Of course, this conclusion does not exclude the possibility
of observing baryon number violation  in
high energy scatterings, because the physical value of $g$ is
finite  and $F(\epsilon g)$ could have  a value close to zero.

Now a speculation on the asymptotic behavior of $H(y)$.
We see in Fig.4 that $H(y)$ oscillates around zero.
It is tempting to assume that $H(y)$  converges to zero
asymptotically.  One may doubt this on the
observation that the amplitude between the origin and the first
 node of $H(y)$
is much smaller than that between  the first and the second node.
However this kind of behavior is expected  to satisfy the
unitarity condition (\ref{r49}); Since the former is negative,
and the integral in (\ref{r42}) is heavily weighted toward the
origin, there must be a large positive region for $H(y)$.
If $H(y)$ indeed converges to zero, the constant $d$
becomes unity and the anomalous cross section is exponentially
suppressed at asymptotic energies. Note that then $g \bar{N}$
also converges to its vacuum value that is unity.  If $ H(y)$
begins to converge at not too large $y$,
higher order terms of $H(y)$ could eventually reveal
the symptom. Thus it is a very interesting problem
to calculate higher order terms of $H(y)$ and
see the functional behavior.

\section{Conclusion}

We gave a new formulation of the anomalous cross
 section from the viewpoint of perturbation theory,
and showed
that order by order our method consistently gives better
high energy behavior for the anomalous cross section.
In our formalism the energy expansion of the
anomalous cross section is an intermediate step toward
the $ \epsilon/ n$ expansion. Using unitarity,
we argued that
under a  plausible condition the
asymptotic form  of  $F$ is  either linear in energy or
a nonnegative constant, and that $F(\epsilon g) > 0$
for all energies. A speculation on the asymptotic behavior of
 $ H(y) $  was made.

\newpage

\appendix{Appendix}

Let us consider the integral in (\ref{r12}),
\be
e^{W(N)} \equiv \int_{ a -i\infty}^{a + i\infty} d z \,
e^{ -z + K(z) + N \ln z}  = \int d z \,e^{ - {\cal F}}.
\ee
We would like to show that in the weak coupling
limit $W$ is exactly given by the
saddle point approximation in (\ref{r13}).
This can be shown conveniently in the Feynman diagram
technique as employed in the proof that the
leading Borel singularity of instanton-induced
amplitudes is determined by the saddle point approximation
\cite{R6}.
To simplify the argument, let us assume $ K(z) $ is
given by the leading term in (\ref{r3}),
\be
K(z) = \frac{1}{2} (3 \epsilon)^{\frac{4}{3}} z^{-\frac{1}{3}}
\equiv C_{o} z^{-\frac{1}{3}}.
\ee
Adding higher order terms  should be trivial.
Expanding ${\cal F}$ about the saddle point $ z(N)$,
\bear
e^{W} &=& \int_{-\infty}^{\infty} d \eta \exp\left\{
-{\cal F}( z(N))
-\frac{1}{2} {\cal F''}(z(N)) \eta^{2} - \sum_{n=3}
\frac{ {\cal F}^{(n)}
(z(N))}{n!} \eta^{n}  \right\} \nonumber \\
     &=& \exp\left\{ - {\cal F}(z(N)) -\frac{1}{2} \ln  \left|
      {\cal F}''(z(N))\right|
     + \sum \left( \mbox{ bubble diagrams} \right) \right\},
      \eear
 where $\eta$-integration is over the real axis. Let us call
 the vertex with $i$ number of legs the $i$-th vertex.
 Now consider a vacuum bubble diagram $B$ with
 $n_{i}$ number of the $i$-th vertex, and $I$ number of
 internal lines. Then
 \be
 \sum_{i} i \cdot n_{i} = 2 I
 \ee
 and
 \bear
 B   &\sim& \frac{ \prod_{i} \left[ {\cal F}^{(i)}
 \left( z(N)\right)
 \right]^{n_{i}}}{
 \left[ {\cal F}^{(2)}\left( z(N)\right)
 \right]^{I}}  \nonumber \\
     &=& N^{-\sum_{i=3} \left( \frac{i}{2} -1 \right)
     n_{i}} \frac{\prod_{i}
     \left( (i-1)!\right)^{n_{i}}  \left( 1 - \frac{
     C_{o} \Gamma(i + 1/3)}{
     N\Gamma(1/3) \Gamma(i)} z(N)^{-\frac{1}{3}}
 \right)^{n_{i}}}{
     \left( 1 - \frac{ 4 C_{o}}{9 N} z(N)^{-\frac{1}{3}}
     \right)^{I}}.
\eear
     Since at the saddle point $ N= \bar{N}$,
     \be
     \frac{C_{o} }{\bar{N}} z(\bar{N})^{-\frac{1}{3}}
 \sim \left(
     \frac{\epsilon }{\bar{N}} \right)
     ^{ \frac{4}{3}} \sim O(1),
     \label{r55}
     \ee
     we have
     \be
     B \sim O \left( N^{-\sum_{i=3} \left( \frac{i}{2} -1
     \right)n_{i}  } \right).
     \ee
     However, to survive the weak coupling limit $B$
     must be at least of $O(N)$, which is impossible.
     Therefore, the corrections to the saddle point
     approximation generates  negligible terms in the
     weak coupling limit.  Similarly from (\ref{r55})
\be
     \ln {\cal F}''(z(N)) + \ln N = O( 1)
\ee  in the weak coupling limit.

\newpage

\parindent 0.0in
\hspace{2.5in} Figure Captions

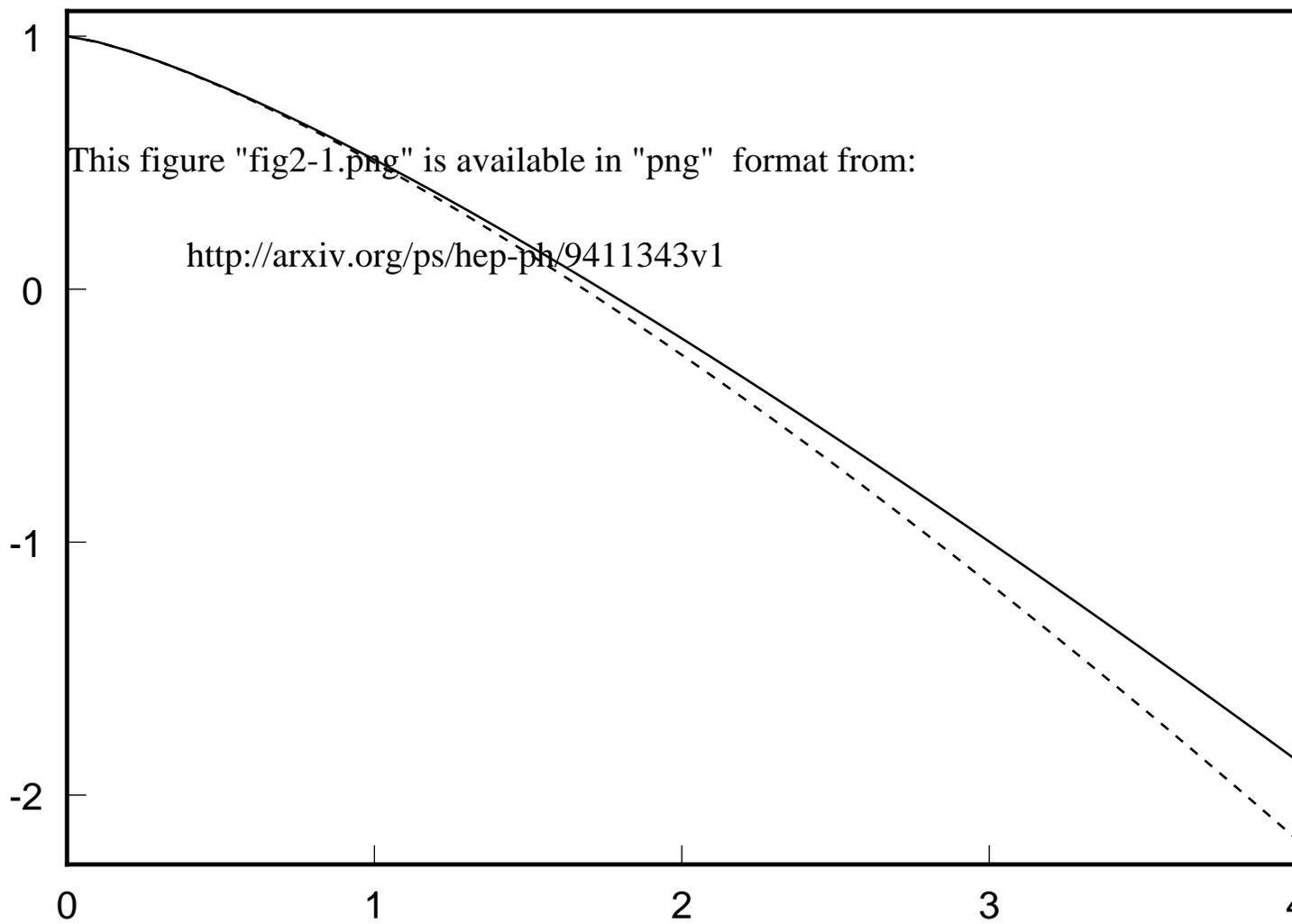
Fig. 1: $F$ and $F_{0}$ versus $ 3
\epsilon g $ to the leading order.
Solid and dashed lines are for $F$ and $F_{0}$ respectively.\

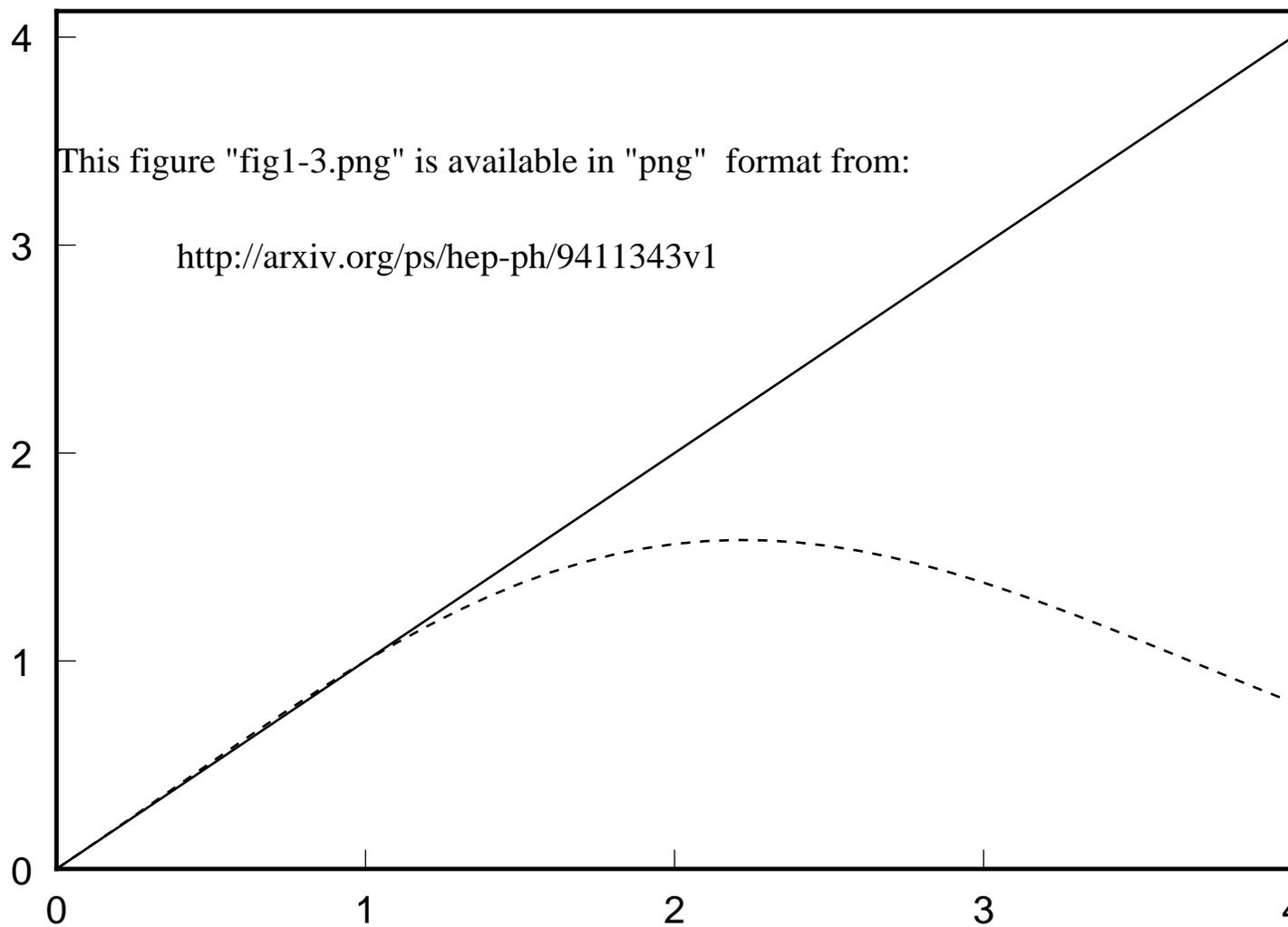
Fig. 2 : The function $ 3 y \exp (-H(y))$ versus $3 y$
 to the second
order. Dashed line is for $ 3 y \exp (-H(y))$. \

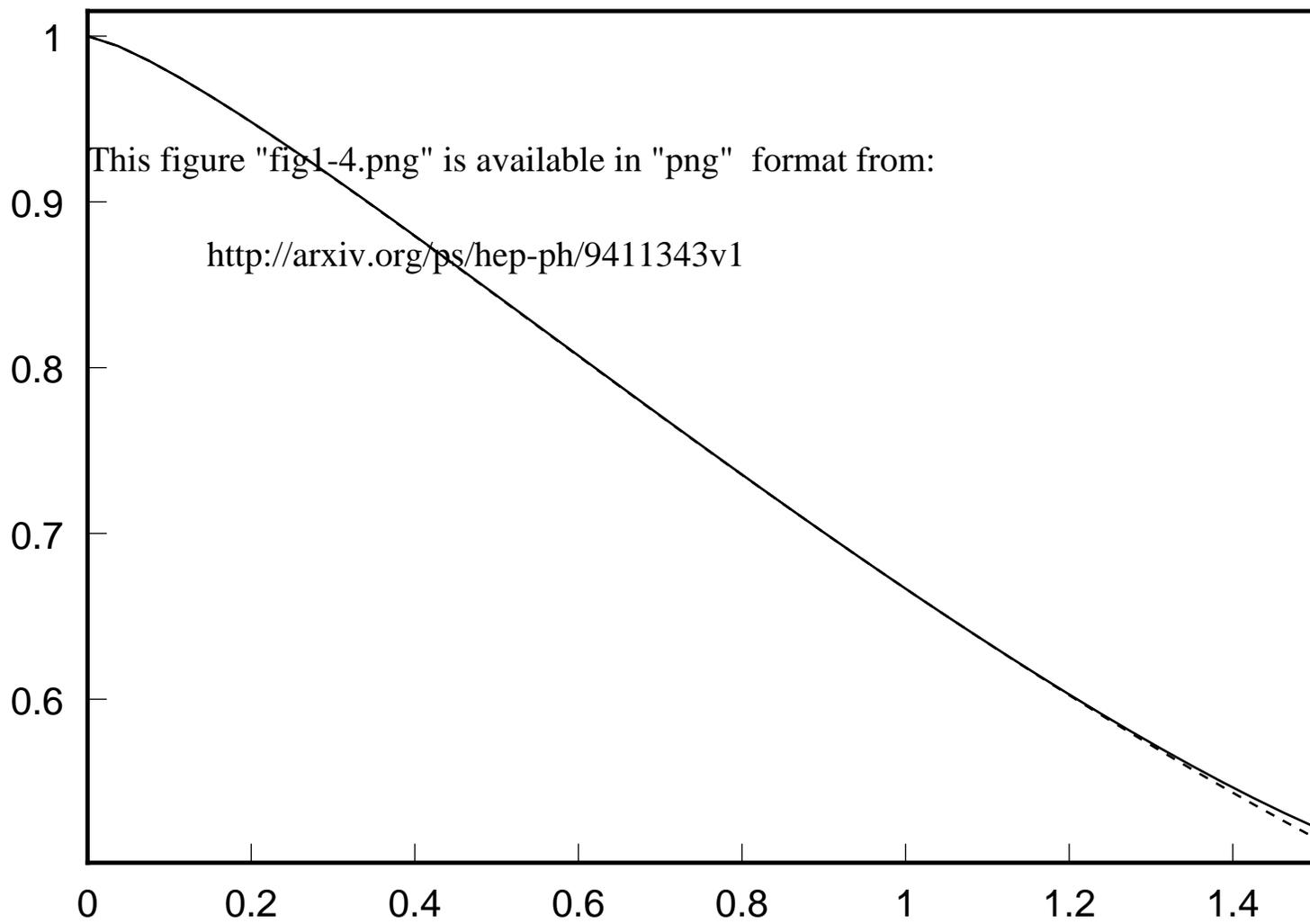
Fig. 3: $F$ and $F_{0}$ versus $ 3 \epsilon g$ to the
second order.
        Solid and dashed lines are for $F$ and $F_{0}$
respectively.\

Fig. 4 a, 4 b:  The functions $ H(y),  3 y \exp (-H(y))$
versus $3 y$
for $\lambda = -1$ and $-0.6$ respectively. Dashed and
 dot-dashed
 lines are for $H(y)$, $ 3 y \exp (-H(y))$  respectively.\

Fig. 5 a, 5 b: $ g \bar{N}, F$ and $F_{0}$
 versus $3 \epsilon g$
to the
third order for $\lambda= -1$ and $-0.6$ respectively.
Note the
large difference
between $F$ and $F_{0}$ at high energies.
For $\lambda = -1$, $g
\bar{N}$ oscillates around unity. Dot-dashed lines
are for $g \bar{N}$, the solid and dotted lines are for
$F$ and $F_{0}$ respectively.

  
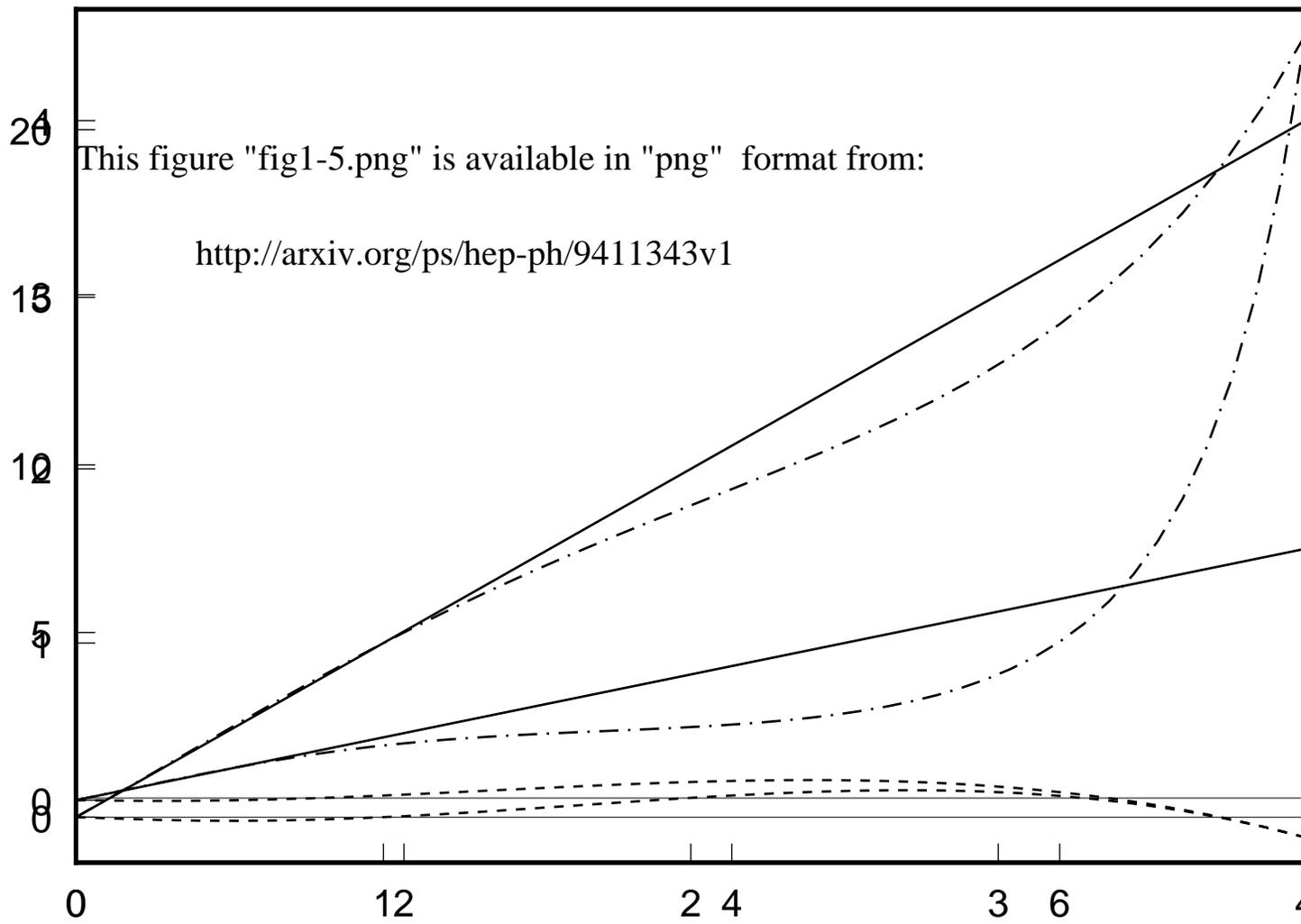
\end{document}